\documentclass[12pt]{article}
\usepackage[utf8]{inputenc}
\usepackage{amsmath,amssymb,amsthm,titlesec,lipsum}
\usepackage[paperwidth=8.5in, paperheight=11in,margin=1in]{geometry}
\usepackage{setspace,lineno}
\usepackage{graphicx} %package to manage images
\graphicspath{ {./Thesis Code/} }
\usepackage{subfig}
\usepackage{listings}

\title{Applications of Random Matrix Theory in Machine Learning and Brain Mapping}
\author{Katrina Lawrence \\ University of Western Ontario}
\date{April 2nd, 2022}

\usepackage[numbers, super]{natbib}
\makeatletter
\renewcommand\@biblabel[1]{#1.}
\makeatother
\titleformat{\section}{\normalsize\bf}{\thesection}{1em}{}
\titlespacing*{\section}{0pt}{6pt}{6pt}

\titleformat{\subsection}[runin]{\normalsize\it}{\thesubsection}{0.5em}{}[---]
\titlespacing*{\subsection}{0pt}{3pt}{3pt}

\begin{document}

\maketitle

\renewcommand{\thefootnote}{} % Remove footnote numbering
\footnotetext{This paper was authored in April 2022 and was published on arXiv in February 2025.}
\renewcommand{\thefootnote}{\arabic{footnote}} % Restore numbering

\begin{abstract}
	% the summary goes here. It won't be 12pt font but don't worry about that.
	Brain mapping analyzes the wavelengths of brain signals and outputs them in a map, which is then analyzed by a radiologist. Introducing Machine Learning (ML) into the brain mapping process reduces the variable of human error in reading such maps and increases efficiency. A key area of interest is determining the correlation between the functional areas of the brain on a voxel (3-dimensional pixel) wise basis. This leads to determining how a brain is functioning and can be used to detect diseases, disabilities, and sicknesses. As such, random noise presents a challenge in consistently determining the actual signals from the scan. This paper discusses how an algorithm created by Random Matrix Theory (RMT) can be used as a tool for ML, as it detects the correlation of the functional areas of the brain. Random matrices are simulated to represent the voxel signal intensity strength for each time interval where a stimulus is presented in an fMRI scan. Using the Marchenko-Pastur law for Wishart Matrices, a result of RMT, it was found that no matter what type of noise was added to the random matrices, the observed eigenvalue distribution of the Wishart Matrices would converge to the theoretical distribution. This means that RMT is robust and has a high test-re-test reliability. These results further indicate that a strong correlation exists between the eigenvalues, and hence the functional regions of the brain. Any eigenvalue that differs significantly from those predicted from RMT may indicate the discovery of a new discrete brain network. 
\end{abstract}

\pagebreak

%\linenumbers

\section*{Introduction}

This paper will explore the applications of Random Matrix Theory (RMT) in machine learning (ML) and brain mapping, which is a subset of medical imaging. ML is a critical area of interest since its self-learning capability is the foundation of Artificial Intelligence (AI). As such, ML is a compilation of algorithms, that when presented with large quantities of data, will be able to continuously improve its given predictions on the data without additional programming nor human intervention. The presence of ML is important in medical imaging since it removes the variable of human error and allows for more information to be processed and analyzed than before, thus, aiding hospitals and research institutions to cut costs and increase efficiency. This allows for a reallocation of resources and can lead to further innovation in the field. In the context of brain mapping, which falls under neuroscience and maps brain wavelengths, machine learning leads to the detection of abnormalities and neurological issues that went previously undetected. Thus, this technique opens the door to diagnosing new neurological disorders and is the foundation towards discovering new treatment methods for patients with said diagnoses.  

The brain is made up of billions of neurons which send and receive information. Specifically, the brain is made up of functionally linked regions that constantly share information with each other \cite{Sandstone}, and a dominant area of research is dedicated to understanding the functional interaction between brain regions, hence the correlation and covariance between the regions. Brain mapping analyzes the wave lengths of the brain and outputs the information into a spatial representation, known as a map, which can then be analyzed to identify irregularities. Traditionally PET scans were used to scan the brain, now functional magnetic resonance imaging (fMRI) is used. Resting state fMRI measures spontaneous fluctuations in the blood-oxygenation-level-dependent (BOLD) signal in gray matter regions \cite{Kim2021} and with this information one can estimate statistical dependencies between gray matter activity arising from different regions\cite{Kim2021}. Another growing area of study is detecting change points in the patterns using maximum eigenvalues from RMT \cite{Kim2021}, where change points are points of inflection where the MRI changes\cite{Wu2017}. 

Noise in the data from fMRI machines cause problems when trying to interpret the data found \cite{Veraart2016}. They can lead to RMT discovering brain networks and functional connectivity that don’t exists \cite{Bansal2021}. They can further lead to RMT discovering brain networks with a poor test-re-test reliability, thus rendering limitations to machine learning in this case as it cannot properly use existing data to formulate predictions when presented with new data. As such, with the convenience of machine learning comes the challenge regarding the evaluation of accuracy of the outputs of the algorithms used and the elimination of the prediction errors that arise. 

Many studies encountered challenges due to small sample sizes, which lead to ill-conditioned problems, such as the discovery of spurious brain networks, or the lack of discovery when one exists. In brain mapping, there are tens of thousands of regions of interest of the brain, known as voxels, yet in many cases there are only hundreds of brain scans \cite{Wernick2010}. It was also found that initial studies ran under the assumption that the functional activity (brain activity in specific regions) remained stationary throughout the scan, yet recent research sheds light on the fact that the functional regions do indeed fluctuate during the time of the scan, hence, the functional activity is dynamic \cite{Kim2021}, thus some previous analysis techniques will no longer yield accurate results. 
RMT is a combination of linear algebra and probability theory. RMT allows one to take multi-dimensional randomly generated data and project it into 2 dimensions or 3 dimensions to be able to analyze and interpret the distribution of such data and the covariance between given categories. Two main aspects of RMT are multi-variate statistics and principal component analysis, which are the key drivers of ML, as ML is based on Big Data. This phenomenon can be explained by the fact that the accuracy of the results of these two methods increase with respect to the quantity of data provided. ML is broken down into two segments. The first being, the development of algorithms that quantify relationships within existing data. The second, is the use of these identified patterns to make predictions based on new data \cite{Wernick2010}. Hence, machine learning is used to carry out future interpretations of the brain scans and diagnosis predictions after the radiologist has initially identified the relationship between the first set of data \cite{Erickson2017}. 
This thesis will explore whether RMT  can detect the same brain network, and functional correlations, when more noise is added to the data set. Furthermore, it will determine if  RMT applied to noisy fMRI data generates brain networks with high test re-test reliability \cite{bansal}. Simulations will be carried out with randomly generated data. Addressing this gap is important since it will help improve the reliability of the machine learning and will eventually render the process of brain mapping autonomous. This allows for hospital resources to be redistributed in other areas of importance. Lastly, this should further lead to more accurate diagnosis which will ultimately benefit the patients and help them receive the optimal care that they require.

\section*{Background} % You can change the title to suit your situation

\subsection*{Brain Mapping} % In case you need subheadings 

The brain is made up of millions of cells called neurons, which send and receive signals. The messages that are being sent to and from all parts of the body are electrical impulses, and these impulses are what create brain waves \cite{Sandstone}. In particular, the brain is made up of functionally linked regions that constantly share information with each other. Brain mapping analyzes the wave lengths of the brain and outputs the information into a spatial representation, known as a map, which can then be analyzed to identify irregularities. The brain is always active, and thus will continuously send and receive signals even in the lack of the presence of a stimulus. This is known as the baseline level. As such, brain activity must be monitored in time intervals to determine the baseline activation, and the activation in the presence of a stimulus \cite{Sandstone}. These data, which has been transformed into a visual map, breaks down the brain activity into the various regions of interest of the brain, known as voxels \cite{Sandstone}. 
Brain scans are predominately done by electroencephalogram (EEG) scans, where a cap is placed over the head which captures the electrical impulses of the brain, and by functional magnetic resonance imaging (fMRI) scans. EEG scans are used to identify the type of brain wave that is being emitted and is typically faster and more affordable method then fMRI scans. However, EEG scans do not provide information on the localization of the source of the brain signals i.e., functional information, as an fMRI scan would \cite{Menon2005}. Thus, the EEG scan will often be used to complement the fMRI scan and this paper will place an emphasis on the use and analysis of fMRI scans. 

fMRI scans measure the functional changes in the brain, known as brain activity, by detecting changes of blood flow in the various regions of the brain (magnetic-resonance.org). This method is based on the fact that cerebral blood flow and neuronal activation are coupled \cite{Logothetis2001}, which means if a region of the brain is activated or in use, then the blood flow to that area increases. Visually, this is seen as a printed grey image of the brain, with regions highlighted in various shades of a colour to indicate the activated regions of the brain with varying degrees of magnitude of activation, see figure 1.
\begin{figure}[htbp]
\centerline{\includegraphics[scale=.75]{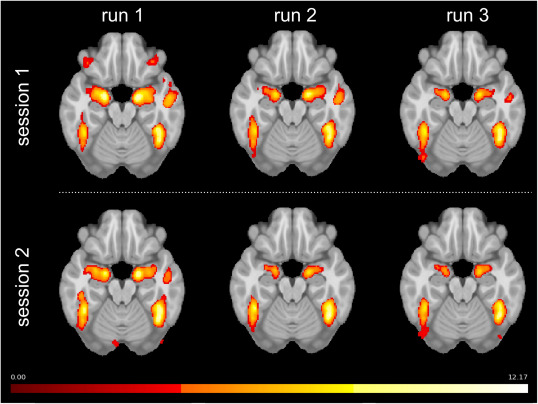}}
\caption{fMRI Brain Scan}
\label{fig}
\end{figure}
To achieve these results, fMRI scans typically use the blood-oxygen-level dependent (BOLD) contrast to detect and then map the cerebral neural activity by identifying and subsequently imaging the change in blood flow related to the energy use by brain cells \cite{Huettel2009}. When an area of the brain becomes activated, the oxygen-rich blood displaces the oxygen depleted blood. However, this process does not happen instantly, it takes approximately 2 seconds on average and peaks after 4 to 6 seconds before returning to baseline or original levels. The fMRI scans are able to capture the difference in the oxygenated and non-oxygenated blood, thus highlighting the activated regions. The deoxygenated blood is more magnetic than the oxygenated blood, which is essentially non-magnetic, and as such, the magnetic resonant (MR) signals from the oxygenated blood interfere less with the MR signals from the deoxygenated blood \cite{Huettel2009}. 

For reference, brain maps that have a strong and evident contrast between areas of interest and nearby locations are known to have high spatial resolution. The higher the spatial resolution of a map, the easier it is to read, and it typically yields more accurate results. The spatial resolution is measured in voxels, where a voxel is a region of interest of the brain and can be thought of as a three-dimensional pixel, and an individual point in space on a three-dimensional, regular matrix \cite{smith2004}.
\begin{figure}[htbp]
\centerline{\includegraphics[scale=.75]{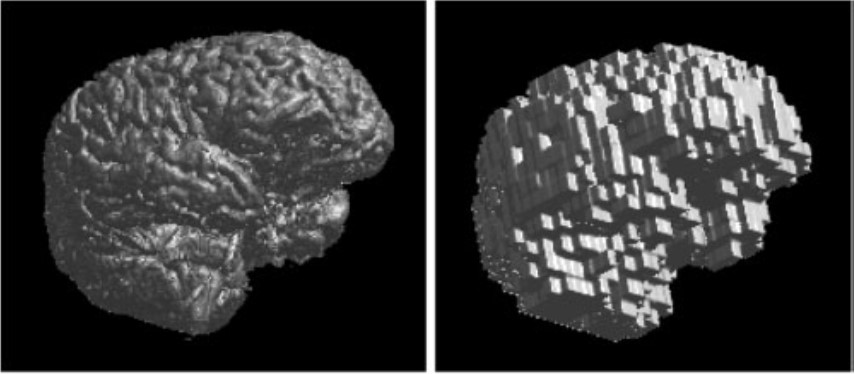}}
\caption{Spatial resolution comparison of brain images}
\label{fig}
\end{figure}
Specifically, in the setting on of a brain scan, a voxel’s dimensions are set by the slice thickness, the area of a slice, and the grid imposed on the slice by the scanning process \cite{Sharoh2019}. A smaller rectangular cuboid leads to a higher resolution scan. See figure 2 for a visual representation of the brain in made from larger voxels (on the right) and smaller voxels (on the left)\cite{smith2004}. As such, full brain scans, which search for more general information, will use larger voxels compared to scans that target a specific region, which will use smaller voxels \cite{Sharoh2019}. A voxel’s response to a stimulus over time is called its time course. During BOLD fMRI scans, a stimulus is presented, and the intensity of the brain signal per voxel is measured per time course. Resting state fMRI scans also exist, they are also known as a taskless fMRI, and they map the brain when no stimuli is presented and thus, they often act as a baseline for BOLD fMRI scans. 

A common challenge in analyzing fMRI scans is removing unwanted noise and leaving just a signal. Unwanted signals are referred to as noise, and often times, the noise can be as large as the signal itself \cite{Huettel2009}, which leads to radiologists presenting stimuli various times to the subject to generate more accurate results during the fMRI scan. The noise comes from various sources, is random, and is challenging to predict. There are five main sources of noise:  system, thermal, physiological, random neural activity and differences in both mental strategies and behaviour across people and across tasks within a person \cite{Huettel2009}. RMT is thus implemented as a tool to alleviate the challenge imposed by noise, by consistently detecting patterns in the brain despite varying levels of random noise.

\subsection*{Random Matrix Theory} 

RMT was introduced by John Wishart in 1928. He is credited with derived Wishart’s Distribution which is generalized product-moment distribution and is fundamental to multivariable statistical analysis, a key component to Random Matrix Theory \cite{Livan2017}.

RMT is now widely used and has found applications in mathematical physics, quantum gravity, engineering, signal processing, mathematical finance, medical imaging, and artificial intelligence, as examples. Random Matrix Theory is a combination of linear algebra and probability theory. This theory allows for multi-dimensional data to be taken and then projected into 2-dimensions or 3-dimensions. This, in turn, enables an interpretation of the distribution of such data to be made and provides an understanding of the covariance between given categories or fields, by looking at the eigenvectors and corresponding eigenvalues of respective random matrices \cite{Livan2017}. RMT is especially known for its remarkable universality, which can be portrayed through the fact that eigenvalue correlations do not depend on the probability distribution \cite{Forrester2003}.

An important aspect of RMT is Multivariable Statistics and Principal Component Analysis. A new space of matrices will be considered, where x1, ... ,xp are random variables.  The covariance of these random variables are of interest. A mean of zero will be assumed, such that 
\begin{equation}
	\ E(x_i) = 0
\end{equation}
and
\begin{equation}
	\ Cov(x_i, x_j) = E(x_i, x_j) – E(x_i| E(x_j)) = E(x_i, x_j)
\end{equation}
 Next, these entries will be put into a matrix, such that a covariance matrix has be created. This covariance matrix will be denoted as 
 \begin{equation}
	\  X = [E(x_i, x_j)]^N_{ij=1}
\end{equation}
Random samples are required to build this matrix. These matrices have N rows and P columns. 

A Wishart Matrix, denoted as W, is created my multiplying the covariance matrix X by its transpose and dividing by the sample size n. It is expressed in the following way: 
\begin{equation}
	\ W = \frac{1}{N}XX^T. 
\end{equation}
Thus, the result is a P by P (where P is a property of the data collected), covariance symmetric matrix with iid entries (independent, and identically distributed) and as consequence all the eigenvalues are non-negative and real.  This is the Marchenko-Pasture Law for Wishart matrices which uses the maximum eigenvalues to model the distribution of the data and the covariance of the random variables \cite{Livan2017}.  A well-known result in RMT is that the spectrum of such Wishart matrices is given by the Marchenko-Pastur distribution \cite{Livan2017}. This theory states that as N approaches infinity and as P approaches infinity, the N/P ratio will converge, and the sample distribution of eigenvalues will converge to the theoretical distribution \cite{Livan2017} \cite{Bansal2021}. The Marchenko-Pastur law shows the asymptotic behaviour of singular values of large rectangular matrices, and this theoretical distribution can be plotted along with the histogram of the eigenvalue distribution to convey the accuracy or the closeness of the observed data distribution to the theoretical one \cite{Lalley2019}. 
Eigenvalues of the Wishart Matrix which deviate significantly from those predicted by RMT likely represent the nonrandom structure of connectivity from brain networks \cite{Bansal2021}. Large deviations from the theoretical distribution my indicate the discovery of a discrete brain network. The eigenvectors associated with the eigenvalues that deviated from the predicted distribution are used in the prediction of such networks \cite{Bansal2021}.

\section*{Methodology}

Various types of clinical fMRI trials are done to identify and analyze correlations between the functional regions of the brain and draw conclusions based on different areas of research. Researchers aim to discover which regions of the brain become activated, and at what magnitude, when a stimulus is applied to the subject. These signal correlations can highlight one’s comprehension on a topic, how their brain is functioning, as well as detecting the presence abnormalities such as a tumor or a prior stroke.

The data that is publicly provided only includes the subjects’ reaction time to the stimuli and not the intensity of the brain signal per voxel. Thus, due to limitations of publicly available information on brain voxel signal intensity, large amounts of data were randomly simulated instead and were inputted in a mathematical framework using RMT to be analyzed. This framework is universal and works for any type of fMRI data given a sufficiently large sample size. The randomly simulated data represents the voxel signal intensity, and simulations of the data are to be carried out using Matlab. 

The random matrices have P columns and N rows. P is the number of properties being examined; thus each column represents a given brain voxel and the inputted value is the brain signal intensity per voxel. N is the number of instances per property; thus, each row represents the time interval over which a stimulus was introduced to a subject during an fMRI scan. As such, the aim is to observe which brain voxels light up during each time interval when a new stimulus is presented and determine the correlation between the activated voxels (Livan, \cite{Bansal2021}. 
Once the random matrices are generated, they are converted into Wishart Matrices from which the eigenvalue distribution is calculated and plotted against the theoretical Marchenko-Pastur distribution. Numerous simulations were ran to test the test-re-test reliability of the Marchenko-Pastur Law for Wishart Matrices, and to determine the overall pattern of the data and the correlation between the voxels at various time intervals. For each simulation, either the random distribution was changed, the size of the matrix was changed, or the level of random noise added to the random matrices was changed. The error between the theoretical distribution and the observed eigenvalue distribution were calculated. See appendix A for the code that was used to run the simulations. 

To understand how a fMRI clinical trial is done, the below example will be provided. A study was done  to determine the reading comprehension level of scientific texts of 52 adults of varying ages, and academic backgrounds, and to determine the neuro-cognitive mechanisms present for successful comprehension \cite{Li2019}. The adults were given five scientific texts to read while they were in the fMRI machine and wore a device which tracked their eye movement. While the participants are in the fMRI machine, they are reading the text. Each phrase is up for 8 seconds and automatically moves to the next one from there. If they are finished reading before the 8 seconds is up, then they can click a button to move to the next phrase. At the end of the section, they are given a 10 question multiple-choice quiz to test their comprehension of the scientific text. While the participants read the text, an eye tracking scanner is on them, and their eye-tracking data is recorded. All the while, their brain activity (BOLD signals) is being monitored and is being displayed on a stimulus display screen (Figure 5 from Reading Brain Methods L1 Adults). To collect the data, they measured the duration of a fixation (a word) for each word in the sentence. Before the onset of the first sentence, a cross fixation is presented on the screen for 6000ms, which is enough time on average for the BOLD signal to return to baseline. Between each phrase, the cross fixation is displayed for 500ms \cite{Li2019}. The voxel size for both the resting state MRI data and fMRI data during reading was 3mm x 3mm x 4mm. The voxel signal intensity data from this trial would be used to create a Wishart Matrix, where each brain voxel would be a column and the intensity for each time course would make the rows of the matrix. The same procedure as done with the random matrices will be carried out to determine the correlation between the functional areas in the brain and if a new discrete brain networks.

\section*{Results}

Random matrices following a normal, uniform and Weibull distribution were generated. Each of these matrices were generated three times with the following sizes: 400 by 700, 1600 by 2800, and 3200 by 5600, such that the ratio N/P (rows/columns) remained constant. 
It was found that no matter what distribution is used to generate the random matrices, which are then converted into Wishart Matrices, the eigenvalue distribution follows the Marchenko-Pastur Distribution. It was observed that as the matrix size increased, all the while keeping the N/P ratio constant, the eigenvalue distribution more closely followed the theoretical distribution. See figure 3, figure 5 and figure 5 for the eigenvalue distributions of the Wishaard matrices generated from the 400 by 700, 1600 by 2800 and 3200 by 5600 sized random matrices following a Weibull distribution. Note that the plots for the normal and uniform distributions followed the same convergence pattern as the Weibull distribution.  

However, the introduction of noise during fMRI scans is very common, and the type of noise is unpredictable and random as it stems from many sources. Physiological fluctuations such as heart rate and breathing cycles can in fact increase correlations and lead to spurious correlations between functional brain areas \cite{Bansal2021}. As such, it is important to first eliminate these factors to get a new baseline of results to work with or be aware of the implications the noise has on the dataset. Conversely, other introductions of noise may decrease the observed correlation of the actual data \cite{Bansal2021} and risks the approaches of RMT not finding correlations and patterns that exist. As such, the previously mentioned simulations were repeated, however, this time with the presence of random noise. These simulations test the robustness of RMT, and its ability to determine the actual data correlation consistently and correctly. Various distributions of random noise were introduced to the random matrices. It was found that despite the introduction of the random noise, the eigenvalue distribution of the Wishart Matrices still followed the Marchenko-Pastur Distribution, hence concluding that a strong correlation between the properties exist. Running these simulations multiple times shows that RMT is well conditioned (i.e. not sensitive to a small change of input) and has a high test-retest reliability which is in line with Bansal’s findings.

This has shown that RMT is a tool for Machine Learning, since in the face of large quantities of data, it yields consistent, well-conditioned results that can be used in further diagnostic analysis. Once the data is inputted into the algorithm, the doctors then have an idea of the correlation between the functional areas of the brain. 

To test the accuracy of the use of RMT the L2 norm was calculated to determine the sum of square error (SSE) between the observed eigenvalue distribution and the theoretical Marchenko-Pastur distribution. The equation is as follows:
\begin{equation}
    \sum_{i=1}^{K} (F-f)^2 = SSE
\end{equation}
Where F is the value of the theoretical Marchenko-Pastur distribution, is the height of the bin of the observed eigenvalue frequency and K is the bin number. K=50 was took.  See figure 6, figure 7 and figure 8 for the absolute value of the error between the theoretical Marchenko-Pastur distribution and the observed eigenvalue distribution per bin for the various sizes of random matrices that follow the Weibull distribution. See figure 9 for a summary table of SSE per random matrix.
As the matrix dimensions increase, the SSE decreases.  Overall, the error is small, thus indicating that this model is robust, and that the eigenvalue distribution converges to the theoretical distribution. Hence, RMT is a reliable tool in determining the correlation of Wishart Matrices given a sufficiently large sample size. 

\begin{figure}%
    \centering
    \subfloat[\centering no noise]{{\includegraphics[width=7.5cm]{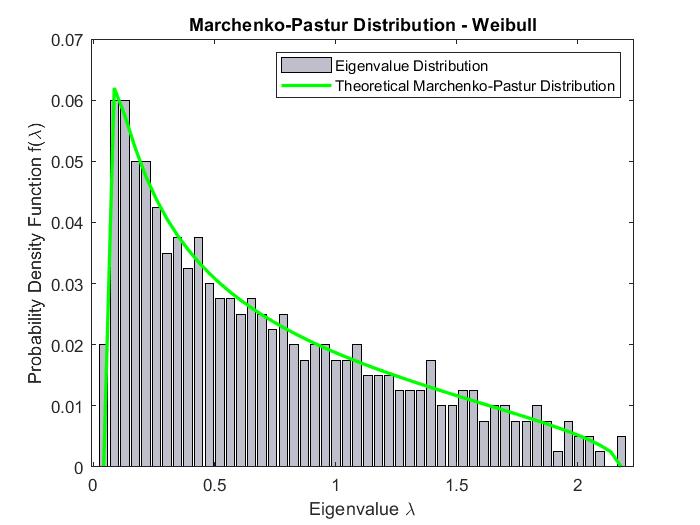} }}%
    \qquad
    \subfloat[\centering with noise]{{\includegraphics[width=7.5cm]{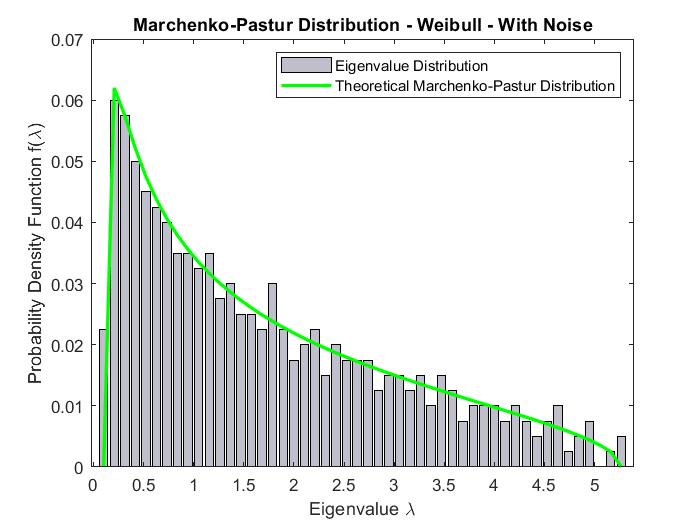} }}%
    \caption{Eigenvalue distributions from 400 by 700 Weibull random matrices}%
    \label{fig:example}%
\end{figure}

\begin{figure}%
    \centering
    \subfloat[\centering no noise]{{\includegraphics[width=7.5cm]{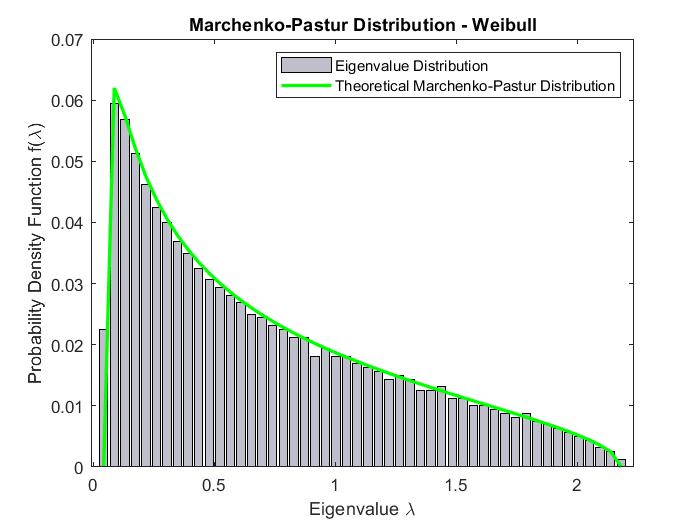} }}%
    \qquad
    \subfloat[\centering with noise]{{\includegraphics[width=7.5cm]{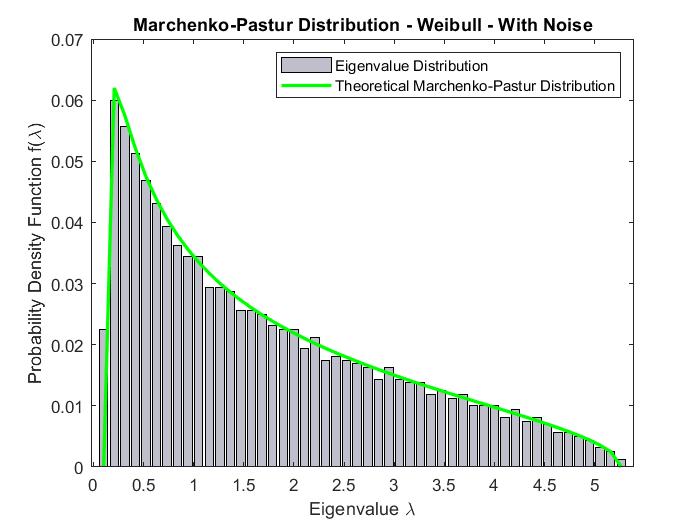} }}%
    \caption{Eigenvalue distributions from 1600 by 2800 Weibull random matrices}%
    \label{fig:example}%
\end{figure}

\begin{figure}%
    \centering
    \subfloat[\centering no noise]{{\includegraphics[width=7.5cm]{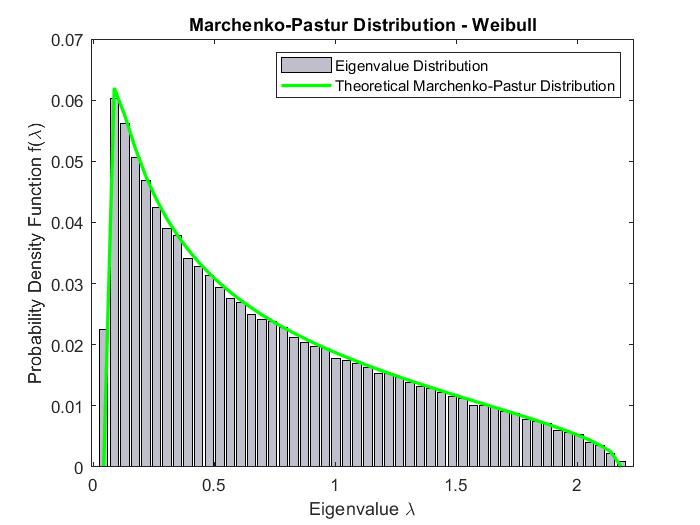} }}%
    \qquad
    \subfloat[\centering with noise]{{\includegraphics[width=7.5cm]{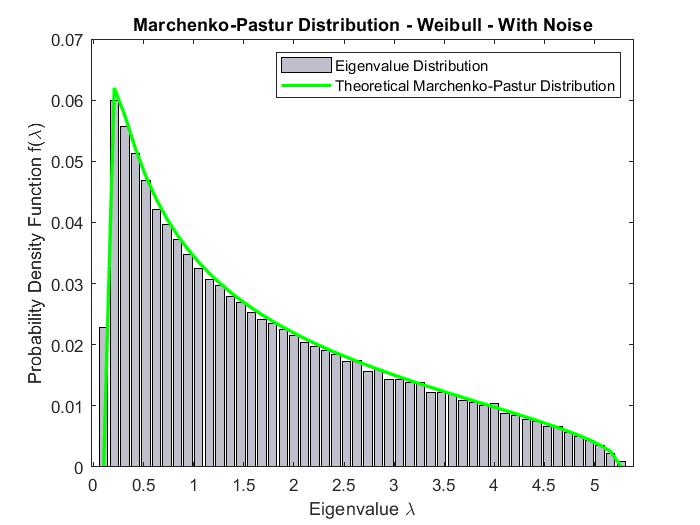} }}%
    \caption{Eigenvalue distributions from 3200 by 5600 Weibull random matrices}%
    \label{fig:example}%
\end{figure}

\begin{figure}%
    \centering
    \subfloat[\centering no noise]{{\includegraphics[width=7.5cm]{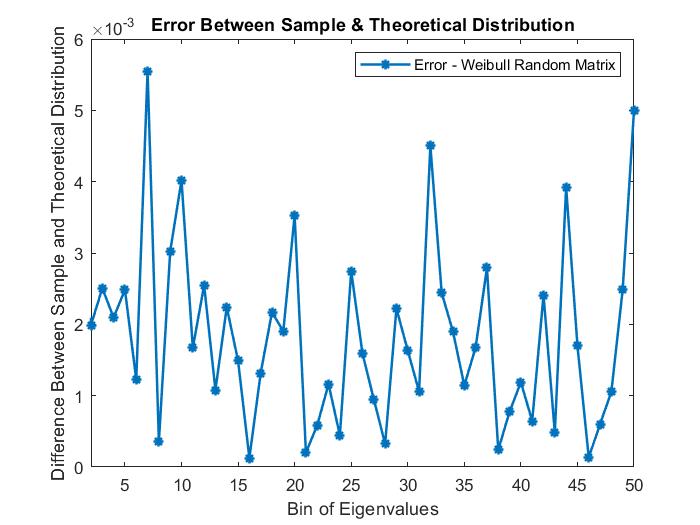} }}%
    \qquad
    \subfloat[\centering with noise]{{\includegraphics[width=7.5cm]{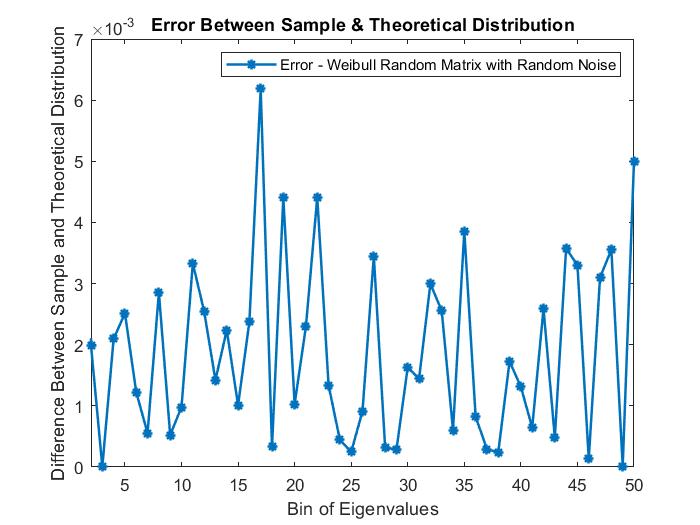} }}%
    \caption{Error between the theoretical Marchenko-Pastur distribution and the eigenvalue distributions from 400 by 700 Weibull random matrices}%
    \label{fig:example}%
\end{figure}

\begin{figure}%
    \centering
    \subfloat[\centering no noise]{{\includegraphics[width=7.5cm]{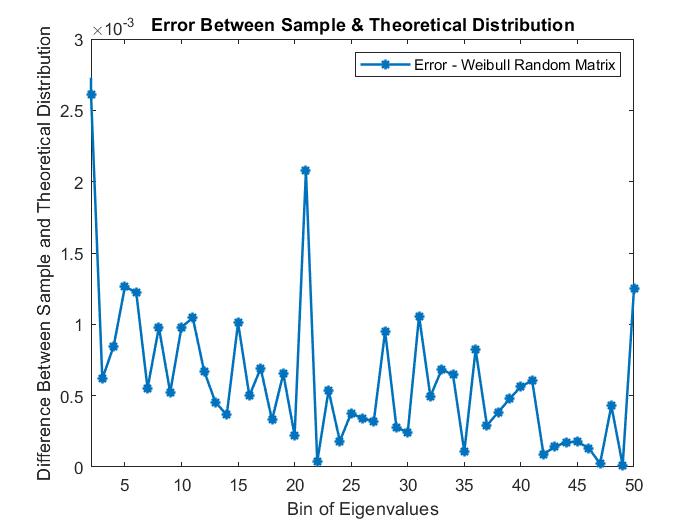} }}%
    \qquad
    \subfloat[\centering with noise]{{\includegraphics[width=7.5cm]{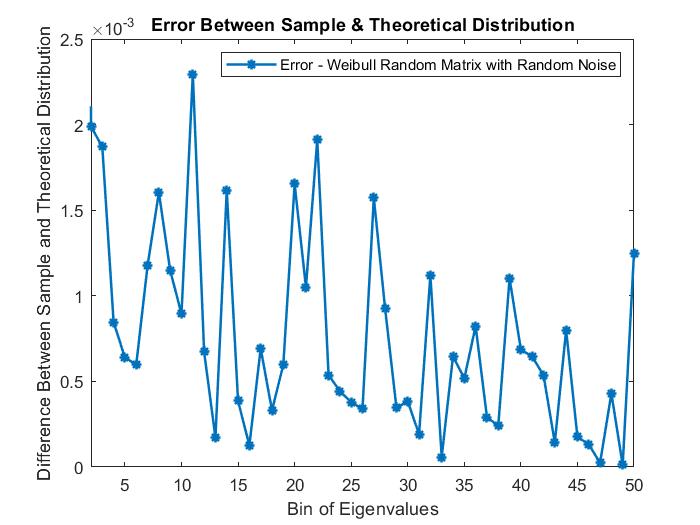} }}%
    \caption{Error between the theoretical Marchenko-Pastur distribution and the  eigenvalue distributions from 1600 by 2800 Weibull random matrices}%
    \label{fig:example}%
\end{figure}

\begin{figure}%
    \centering
    \subfloat[\centering no noise]{{\includegraphics[width=7.5cm]{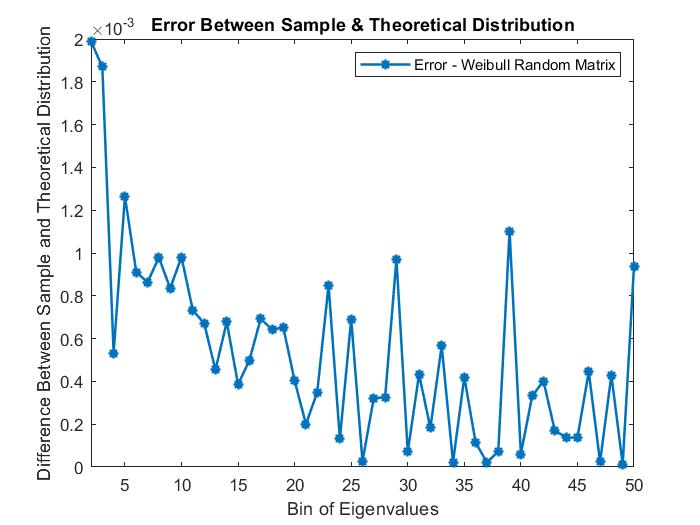} }}%
    \qquad
    \subfloat[\centering with noise]{{\includegraphics[width=7.5cm]{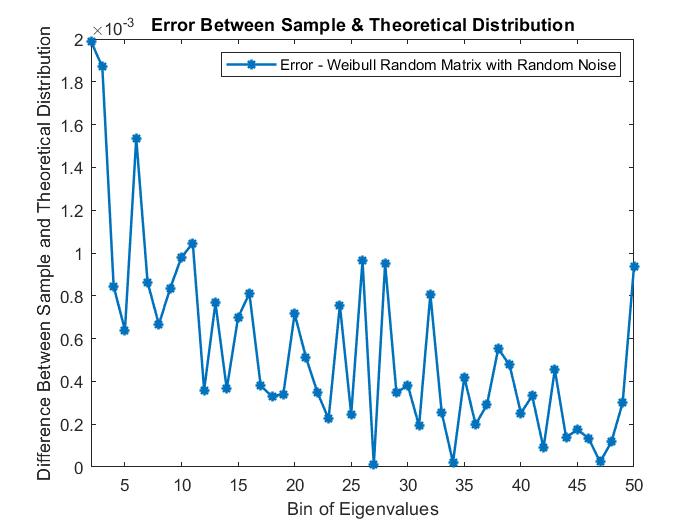} }}%
    \caption{Error between the theoretical Marchenko-Pastur distribution and the eigenvalue distributions from 3200 by 5600 Weibull random matrices}%
    \label{fig:example}%
\end{figure}

\begin{figure}[htbp]
\centerline{\includegraphics[scale=0.8]{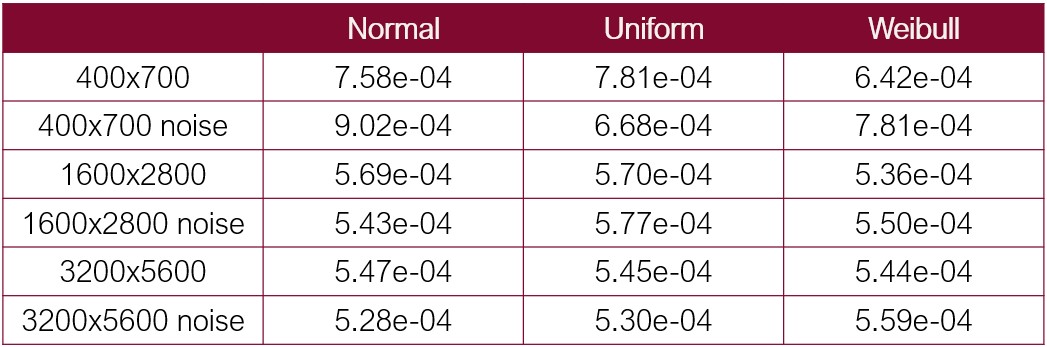}}
\caption{SSE Table per Matrix}
\label{fig}
\end{figure}

\section*{Discussion}

This paper used randomly generated matrices, where each entry was meant to represent the intensity of the brain signal in each voxel over various time courses, to portray how RMT can be used understand correlations, and find patterns despite the presence of noise. It was found that despite the introduction of noise and no matter what probability distribution was used to generate the matrix, the eigenvalue distribution of the Wishart Matrix converged to the theoretical Marchenko-Pastur distribution. This implies that the model is robust and has a high test-re-test reliability. Meaning, this RMT algorithm is not sensitive to changes in initial conditions, and it can consistently determine the correlation between the eigenvalues, hence, the correlation between the functional areas of the brain. 

Given that the brain has a finite number of regions, and that a finite number of time courses will be used when conducting experiments, the possibility exists that the use of theoretical distribution for accessing statistical significance of the brain networks may lead to spurious networks \cite{Bansal2021}. As such, it’s important to use sufficiently large sample sizes, and introduce random noise to the data set to determine if the same patterns continue to emerge, as well as account for correlations of functional regions that intrinsically exists due to physiological fluctuations \cite{Bansal2021}. Bansal found that the RMT method for determining functional correlations and discovering discrete brain networks had a high test-retest reliability when working with real-world datasets\cite{Bansal2021}. Other mathematicians used related approaches to RMT to analyze functional network connectivity from fMRI scans. Liew S and Roshchupkin used independent component analysis (ICA) when the functional network connectivity was highly dynamic, by taking the mean connectivity over time courses \cite{Liew2018}. They further utilized voxel-based morphometry (VBM), where they analyzed the correlation of the signal intensity of the voxels. This was done through generalized linear modeling of the voxel-wise gray matter concentration between different groups \cite{Liew2018}. They found their methods to be accurate and successful. Thus, my findings strengthen the results found from the other mathematicians as it shows that applying another mathematical algorithm to brain mapping yields consistent results in line with what they also discovered. Hence giving the results more validation. 

Looking ahead, brain mapping has a wide spectrum of applications and has specifically made great strides to improving processes and diagnoses in the medical field. Analyzing the intensity of the signals per voxel and the correlation of the voxels across the various functionally linked regions of the brain enables the identification of the regions linked to critical functions such as speaking, moving, sensing, or planning \cite{Vignesh2017}. Furthermore, the brain maps aid in determining how risky brain surgery, as well as assessing how a normal, diseased, or injured brain is functioning. Specifically, brain maps may detect the impact of tumors, strokes, or other head injuries as well as diseases or disabilities like Alzheimer’s, schizophrenia, and Autism \cite{Vignesh2017}. Developing machine learning algorithms, using RMT as a tool, will increase the accuracy and efficiency of brain mapping by decreasing human time and error, thus ultimately leading to greater patient care and medical innovations.

% Appendices if you need them

\renewcommand{\theequation}{A\arabic{equation}}
% redefine the command that creates the equation number.
\renewcommand{\thetable}{A\arabic{table}}
\renewcommand{\thefigure}{A\arabic{figure}}
\setcounter{equation}{0}  % reset counter 
\setcounter{figure}{0}
\setcounter{table}{0}

\section*{Appendix A: Computer Code}

\begin{lstlisting}
The base code that was used to run the simulations is as follows:
clear;
close all;
N = 3200;
T = 5600;
% Ratio of matrix dimentions
c = N/T;

%Need to create a random sample
%x = randn(N,T); %normal distribution
%x = rand(N,T); %uniform distribution
sz = [3200 5600];
x = wblrnd(4,5,sz);

%adding random noise to the matrix
%gaussian distributed noise with mean 0 and variance 1 
%test out different variances
x_wnoise = x + 1*randn(size(x));

%Standard deviation 
s = std(x_wnoise(:));

%spectral matrix
r = x_wnoise*x_wnoise'/T;

%solving for the eigenvalues of r
e = eig(r);
%Probability Density Function
%Number of points for measurement
n = 50;
%Boundaries
a = (s^2)*(1-sqrt(c))^2;
b = (s^2)*(1+sqrt(c))^2;

%making a histogram of eigenvalues of r with 50 bins between a and b
[f,lambda] = hist(e, linspace(a,b,n));
%normalization of the eigenvalues
f = f/sum(f);
%Theoretical pdf formula of M-P law distribution
ft = @(lambda, a,b,c)(1./(2*pi*lambda*c*s^(2))).*sqrt((b-lambda).*(lambda-a));
F = ft(lambda,a,b,c);
%Processing numerical pdf
F = F/sum(F);
F(isnan(F)) = 0;

%Plotting the results
figure(1);
h = bar(lambda, f);
set(h, 'FaceColor', [.75 .75 .8]);
set(h, 'Linewidth', 0.25);
xlabel('Eigenvalue \lambda');
ylabel('Probability Density Function f(\lambda)');
title('Marchenko-Pastur Distribution - Weibull - With Noise');
legend('Eigenvalue Distribution');
emin = min(e);
emax = max(e);
%axis([-1 2*emax 0 max(f)+max(f)/4]);

hold on;
%plotting the smooth curve of the theoretical distribution
plot(lambda, F, 'g', 'Linewidth',2);
legend('Eigenvalue Distribution','Theoretical Marchenko-Pastur Distribution')
hold off;

%creating new array without the max eigenvalue to see if it changes the
%skew in the tails
g = e(e < (emax - 0.0001));

%L2 norm sum((F-f).^2)- gives the sum of square error
SSE = sum((F-f).^2);

%plotting the difference between the bin height and the theoretical
%distribution for each bin - see that it decreases overtime
figure(2)
plot(abs(F-f),'-*','Linewidth',1.5);
xlim([2 50]);
title('Error Between Sample & Theoretical Distribution');
xlabel('Bin of Eigenvalues');
ylabel('Difference Between Sample and Theoretical Distribution');
legend('Error - Weibull Random Matrix with Random Noise');
\end{lstlisting}

\renewcommand{\theequation}{B\arabic{equation}}
% redefine the command that creates the equation number.
\renewcommand{\thetable}{B\arabic{table}}
\renewcommand{\thefigure}{B\arabic{figure}}
\setcounter{equation}{0}  % reset counter 
\setcounter{figure}{0}
\setcounter{table}{0}

\bibliography{mybib} % Your bib file

\begin{thebibliography}{10}
\expandafter\ifx\csname url\endcsname\relax
  \def\url#1{\texttt{#1}}\fi
\expandafter\ifx\csname urlprefix\endcsname\relax\def\urlprefix{URL }\fi
\providecommand{\bibinfo}[2]{#2}
\providecommand{\eprint}[2][]{\url{#2}}

\bibitem{Sandstone}
\bibinfo{title}{What is brain mapping?}
\newblock \emph{\bibinfo{journal}{Sandstone Health}}  (\bibinfo{year}{2021}).

\bibitem{Kim2021}
\bibinfo{author}{Kim, J.}, \bibinfo{author}{Jeong, W.} \& \bibinfo{author}{Chung, C.~K.}
\newblock \bibinfo{title}{Dynamic functional connectivity change-point detection with random matrix theory inference}.
\newblock \emph{\bibinfo{journal}{Frontiers in Neuroscience}} \textbf{\bibinfo{volume}{0}}, \bibinfo{pages}{445} (\bibinfo{year}{2021}).

\bibitem{Wu2017}
\bibinfo{author}{Wu, D.} \emph{et~al.}
\newblock \bibinfo{title}{Mapping the order and pattern of brain structural mri changes using change-point analysis in premanifest huntington's disease}.
\newblock \emph{\bibinfo{journal}{Wiley Periodicals Inc}}  (\bibinfo{year}{2017}).

\bibitem{Veraart2016}
\bibinfo{author}{Veraart, J.}, \bibinfo{author}{Fieremans, E.} \& \bibinfo{author}{Novikov, D.~S.}
\newblock \bibinfo{title}{Diffusion mri noise mapping using random matrix theory}.
\newblock \emph{\bibinfo{journal}{Magnetic Resonance in Medicine}} \textbf{\bibinfo{volume}{76}}, \bibinfo{pages}{1582} (\bibinfo{year}{2016}).

\bibitem{Bansal2021}
\bibinfo{author}{Bansal, R.} \& \bibinfo{author}{Peterson, B.}
\newblock \bibinfo{title}{Use of random matrix theory in the discovery of resting state brain networks | elsevier enhanced reader}.
\newblock \emph{\bibinfo{journal}{Elsevier}} \bibinfo{pages}{69--87} (\bibinfo{year}{2021}).

\bibitem{Wernick2010}
\bibinfo{author}{Wernick, M.~N.}, \bibinfo{author}{Yang, Y.}, \bibinfo{author}{Brankov, J.~G.}, \bibinfo{author}{Yourganov, G.} \& \bibinfo{author}{Strother, S.~C.}
\newblock \bibinfo{title}{Machine learning in medical imaging}.
\newblock \emph{\bibinfo{journal}{IEEE Signal Process Magazine}}  (\bibinfo{year}{2010}).

\bibitem{Erickson2017}
\bibinfo{author}{Erickson, B.}, \bibinfo{author}{Korfiatis, P.}, \bibinfo{author}{Akkus, Z.} \& \bibinfo{author}{Kline, T.}
\newblock \bibinfo{title}{Machine learning for medical imaging 1}.
\newblock \emph{\bibinfo{journal}{RadioGraphics}}  (\bibinfo{year}{2017}).

\bibitem{Menon2005}
\bibinfo{author}{Menon, V.} \& \bibinfo{author}{Crottaz-Herbette, S.}
\newblock \bibinfo{title}{Combined eeg and fmri studies of human brain function}.
\newblock \emph{\bibinfo{journal}{Standford University School of Medicine}}  (\bibinfo{year}{2005}).

\bibitem{Logothetis2001}
\bibinfo{author}{Logothetis, N.~K.}, \bibinfo{author}{Pauls, J.}, \bibinfo{author}{Augath, M.}, \bibinfo{author}{Trinath, T.} \& \bibinfo{author}{Oeltermann, A.}
\newblock \bibinfo{title}{Neurophysiological investigation of the basis of the fmri signal}.
\newblock \emph{\bibinfo{journal}{Max Planck Institute for Biological Cybernetics}}  (\bibinfo{year}{2001}).
\newblock \urlprefix\url{www.nature.com}.

\bibitem{Huettel2009}
\bibinfo{author}{Huettel, S.}, \bibinfo{author}{Song, A.} \& \bibinfo{author}{McCarthy, G.}
\newblock \emph{\bibinfo{title}{Functional Magnetic Resonance Imaging}}, vol.~\bibinfo{volume}{84} (\bibinfo{publisher}{Yale Journal of Biology and Medicine}, \bibinfo{year}{2009}).

\bibitem{smith2004}
\bibinfo{author}{Smith, S.~M.}
\newblock \bibinfo{title}{Overview of fmri analysis}.
\newblock \emph{\bibinfo{journal}{The British Journal of Radiology}} .

\bibitem{Sharoh2019}
\bibinfo{author}{Sharoh, D.} \emph{et~al.}
\newblock \emph{\bibinfo{title}{Laminar Specific fMRI Reveals Directed Interactions in Distributed Networks During Language Processing}}, vol.~\bibinfo{volume}{15} (\bibinfo{year}{2019}).

\bibitem{Livan2017}
\bibinfo{author}{Livan, G.}, \bibinfo{author}{Novaes, M.} \& \bibinfo{author}{Vivo, P.}
\newblock \emph{\bibinfo{title}{Introduction to Random Matrices Theory and Practice}} (\bibinfo{publisher}{Springer}, \bibinfo{year}{2017}).

\bibitem{Forrester2003}
\bibinfo{author}{Forrester, P.~J.}, \bibinfo{author}{Snaith, N.~C.} \& \bibinfo{author}{Verbaarschot, J. J.~M.}
\newblock \bibinfo{title}{Developments in random matrix theory}.
\newblock \emph{\bibinfo{journal}{University of Melbourne}}  (\bibinfo{year}{2003}).

\bibitem{Lalley2019}
\bibinfo{author}{Lalley, S.}
\newblock \bibinfo{title}{Random matrices: Wigner and marchenko-pastur theorems 1 wigner's theorem}.
\newblock \emph{\bibinfo{journal}{University of Chicago}}  (\bibinfo{year}{2019}).

\bibitem{Li2019}
\bibinfo{author}{Li, P.} \emph{et~al.}
\newblock \bibinfo{title}{The reading brain project methods for data collection (l1 adults)}.
\newblock \emph{\bibinfo{journal}{Brain, Language, and Computation Labratory}}  (\bibinfo{year}{2019}).

\bibitem{Liew2018}
\bibinfo{author}{Gazula1, H.} \emph{et~al.}
\newblock \bibinfo{title}{Decentralized analysis of brain imaging data: Voxel-based morphometry and dynamic functional network connectivity}.
\newblock \emph{\bibinfo{journal}{Frontiers in Neuroinformatics}} \textbf{\bibinfo{volume}{1}}, \bibinfo{pages}{55} (\bibinfo{year}{2018}).
\newblock \urlprefix\url{www.frontiersin.org}.

\bibitem{Vignesh2017}
\bibinfo{author}{Subbaraju, V.}, \bibinfo{author}{Sundaram, S.} \& \bibinfo{author}{Narasimhan, S.}
\newblock \bibinfo{title}{Identification of lateralized compensatory neural activities within the social brain due to autism spectrum disorder in adolescent males}.
\newblock \emph{\bibinfo{journal}{European Journal of Neuroscience}}  (\bibinfo{year}{2017}).

\end{thebibliography}
\end{document}